# Observation of acoustic spatiotemporal vortices

Hongliang Zhang[1#], Yeyang Sun[1#], Junyi Huang[1#], Bingjun Wu,
Zhaoju Yang[1*], Konstantin Y. Bliokh[2,3,4], and Zhichao Ruan[1**]

[1]*Interdisciplinary Center of Quantum Information, State Key Laboratory of Modern Optical Instrumentation, and Zhejiang Province Key Laboratory of Quantum Technology and Device, Department of Physics, Zhejiang University, Hangzhou 310027, China*
[2]*Theoretical Quantum Physics Laboratory, Cluster for Pioneering Research, RIKEN, Wako-shi, Saitama 351-0198, Japan*
[3]*Centre of Excellence ENSEMBLE3 Sp. z o.o., 01-919 Warsaw, Poland*
[4]*Donostia International Physics Center (DIPC), Donostia-San Sebastián 20018, Spain*

[#] These authors equal contributions to this work
[*] zhaojuyang@zju.edu.cn
[**] zhichao@zju.edu.cn

**Abstract**

Vortices in fluids and gases have piqued the interest of human for centuries. Development of classical-wave physics and quantum mechanics highlighted wave vortices characterized by phase singularities and topological charges. In particular, vortex beams have found numerous applications in modern optics and other areas. Recently, optical *spatiotemporal* vortex states exhibiting the phase singularity both in space and time have been reported. Here, we report the first generation of *acoustic* spatiotemporal vortex pulses. We utilize an acoustic meta-grating with mirror-symmetry breaking as the spatiotemporal vortex generator. In the momentum−frequency domain, we unravel that the transmission spectrum functions exhibit a topological phase transition where the vortices with opposite topological charges are created or annihilated in pairs. Furthermore, with the topological textures of the nodal lines, these vortices are robust and exploited to generate spatiotemporal vortex pulse against structural perturbations and disorder. Our work paves the way for studies and applications of spatiotemporal structured waves in acoustics and other wave systems.

***Introduction.*** Wave vortices, i.e., structures with the wavefield intensity vanishing in the center and the phase winding around, are of enormous importance for various areas of physics. They are essential parts of almost any structured waves: atomic orbitals and superfluids in quantum mechanics, complex wave interference from ocean waves to nanophotonics and metamaterials, etc. Cylindrical vortex beams have been generated and found applications in electromagnetic [1-8], sound [9-17], elastic [18], electron [19-21], neutron [22], and atom [23] waves. Such states contain on-axis vortex lines and carry intrinsic orbital angular momentum (OAM) along their propagation direction.

Recently, there was a great rise of interest in *spatiotemporal vortex pulses* (STVPs), which are generalizations of usual 'spatial' vortex states to the space-time domain and the OAM tilted with respect to the propagation direction [24-40]. This conforms with the rapidly growing field of space-time structured waves allowing manipulation in both spatial and temporal degrees of freedom [41,42].In the simplest case, STVPs are flying doughnut-shaped pulses with the vortex line and OAM orthogonal to their propagation direction. Until now, such states have been generated only in optics, although theoretically these were also discussed for quantum matter and acoustic waves [33].

Here, we report the first generation of acoustic STVPs for *sound* waves in air. Our STVP generator is based on a meta-grating with spatial mirror symmetry breaking, which can be further controlled by a synthetic parameter. Through mapping the transmission spectra of the meta-grating as a function of the synthetic parameter, we show that there exist vortices in the momentum−frequency domains created and annihilated together in pairs at a critical point by mirror-symmetry breaking. In contrast to diffraction gratings with fork-like dislocations, which are used for the generation of spatial vortex beams in the first and higher orders of diffraction [1,3,4,19], this method uses the zeroth-order transmitted field and realizes simultaneous control in both spatial and temporal domains. Importantly, similar to the topological textures for electronic and optical systems [43,44], these vortices associated with the nodal lines are robust and exploited to generate STVPs with the topological protection against structural disorder. Our results open the avenue for spatiotemporal vortex generation and applications in acoustics and other areas of wave physics [7,11,32,45-49].

***Breaking spatial mirror symmetry for generating STVPs.*** Our STVP generator is based on spatial mirror symmetry breaking [32]. Figure 1a schematically displays the spatial-symmetry analysis of the spatiotemporal vortex generation [the detailed geometry in Supplementary Material (SM) Sec. I]. Here, a spatiotemporal Gaussian pulse impinges on the structure along $x = 0$ (indicated by the white dashed line). Without loss of generality, assume that a meta-grating exists mirror symmetry about the plane $x = 0$, the phase distribution of the transmitted pulse must also be symmetric about the mirror plane, and thus there is no phase singularity. Therefore, the necessary condition for generating STVPs with nonzero winding numbers is the mirror symmetry breaking, which provides an asymmetry modulation for sound in both spatial and temporal domains simultaneously.

To realize the asymmetric spatiotemporal modulation, we design the meta-grating (Fig. 1a) with a unit cell consisting of four air blocks with different sizes as indicated by dashed boxes in SM Fig. S1. In the meta-grating, all cells are connected with a middle air channel (yellow areas in Fig. 1a). Initially, all the four air blocks are symmetric about the axis of $x = 0$. To break the spatial mirror symmetry, we introduce a synthetic parameter $\eta$, which describes the different $x$-directional shifts from the center $\delta x_i = A_i \eta$, where $A_i$ is the shifting ratio given in Supplementary Table S1 for the $i$-th block and $i = 1, 2, 3, 4$. Thus, $\eta$ controls the degree of the mirror asymmetry of the grating.

This asymmetric modulation stemming from mirror symmetry breaking can be illustrated by the transmission spectrum function $T(k_x, \omega)$ between the incident and transmitted waves, where $\omega$ is the angular frequency of a plane wave, and $k_x$ is the wavevector along the structure interface. Here only the zeroth order diffraction is considered in the operating-frequency range. For the mirror symmetric case of $\eta = 0$, as expected, the transmission spectrum function $T(k_x, \omega)$ exhibits symmetric spectrum about $k_x = 0$ for both the phase and the amplitude distributions (Fig. 1b). On the other hand, by breaking the mirror symmetry, the case of $\eta = 0.5$ shows that the transmission spectrum function $T(k_x, \omega)$ has two vortices with the winding numbers of $l = +1$ (white circle) and $l = -1$ (black circle), respectively (Fig. 1c). Correspondingly, such two phase singularities coincide with the zero-value transmission at the center of the vortices. Furthermore, for $\eta = 0.7$, the two vortices are further separated with a larger strength of mirror symmetry breaking (Fig. 1d).

The vortices associated with the transmission spectrum function are analogous to the fundamental 'charges' in the $k_x - \omega$ domain, because they are always created or annihilated together in pairs of opposite charges. To clearly illustrate the creation and annihilation, Fig. 1e shows the parameter $\eta$ spectrum, which is in the form of nodal lines in the extended dimension with the asymmetry parameter $\eta$. When a plane with a fixed value of $\eta$ has intersections with the nodal lines, there must be two vortices with opposite handedness appearing in the cross-section plane. Therefore, the total topological charge of the vortices is a conserved quantity of zero. Through numerical simulations, we find that the critical value of $\eta$ is $\eta_c = 0.40$ for our designed meta-grating.

To experimentally demonstrate the topological phase transition, we fabricate three meta-gratings (SM Fig. S3) and measure the transmission spectrum function (SM Sec. II). Figure 2a-c show the measured phase distributions of the transmission spectrum function for $\eta = 0, 0.5, 0.7$, and the corresponding amplitude distributions are shown in Fig. 2d-f. For the mirror symmetric case (Fig. 2a and d), there is no phase singularity in the measured phase distribution of the transmission spectrum function. By breaking the mirror symmetry with $\eta = 0.5$, there are two vortices created with the winding numbers of $+1$ and $-1$, marked by the white and black circles in Fig. 2b, and hence two zero-value transmissions occur at $\omega/2\pi = 8.31\ kHz$ and $\omega/2\pi = 8.23\ kHz$ in Fig. 2e, respectively. For the case of $\eta = 0.7$ as shown in Fig. 2c and f, two vortices with opposite winding numbers are further separated at $\omega/2\pi = 8.31\ kHz$ and $\omega/2\pi = 8.04\ kHz$. The measured results agree that the topological phase transition appears at the critical point of $\eta_c = 0.40$. The creation of the vortices confirms that the asymmetric modulation by breaking mirror symmetry indeed induces the topological charges in the $(k_x, \omega)$ domain.

***Topological protected generation of STVP.*** Owing to the asymmetric modulation, the phase singularity of the transmission spectrum function located at $(\omega_0, k_{0x})$ in the $(k_x, \omega)$ domain can be directly transferred into the spatiotemporal domain. Considering an incident Gaussian wave pulse with the central angular frequency $\omega_0$ and transverse wavevector component $k_{0x}$, the transmitted wave packet can be determined to be a STVP with a nonzero winding number, but opposite to that of the vortex in the $(k_x, \omega)$ domain (see SM Sec. IV). We also note that the nodal line in the space of $\omega, k_x, \eta$ is mathematically analog to many topological textures in other nodal-line topological physical systems [43,44]. Because of the topological robustness of the nodal lines, the corresponding vortices with the winding number $l = \pm 1$ are stable. When the vortices are distanced from the critical points of the topological phase transition, small changes to the meta-grating geometry in the real space can be treated as perturbations. The strength of such topological protection for the vortex at $(\omega_0, k_{0x})$ can be evaluated by the distance to its nearest vortex $\Delta = \sqrt{(\omega_n - \omega_0)^2 + v^2(k_{nx} - k_{0x})^2}$, where $\omega_n$ and $k_{nx}$ are the frequency and the wavevector component of the nearest vortex.

We next experimentally demonstrate the topologically protected generation of acoustic STVPs, schematically displayed in Fig. 3a. Here, we choose the meta-grating with the asymmetry parameter $\eta = 1.0$, because such a meta-grating has a relatively strong strength of topological protection up to $\Delta$. We first experimentally investigate the transmission spectrum function (SM Sec. V), which exhibits a vortex at $\omega_0/2\pi = 8.02\ kHz$ and $k_{0x} = 0.01k_0$ ($k_0$ is the wavenumber in air). For $k_{0x} \ll k_0$, therefore, it can be used as for normally-incident pulses. Using an arc-like linear transducer array, with an oscillatory Gaussian envelopes electric signal at a central carrier frequency $\omega_0$, we produce a $z$-propagating Gaussian pulses, which has the full waist of 145 mm and the durations of 3.2 ms and the diffraction Rayleigh range about 383 mm (the detailed experimental setup and the measurement found in SM Sec. VI. and Fig. S6).

We first consider the unperturbed meta-grating as the experimental sample Fig. 3b. We numerically simulate the transmitted pulse envelope at different propagation distances of 64.4mm, 78.8mm, 93.2mm, and 107.6mm, which are separated by $\frac{\lambda}{3}$ from each other ($\lambda$ is the center wavelength of the pulse) as indicated by red dash lines in Fig. 3a, respectively. Here the transmitted acoustic wave $P(x,t) = s_{out}(x,t)e^{-i\omega_0 t}$, where $s_{out}(x,t)$ is the transmitted pulse envelope. The simulation results are depicted in Fig. 3d-g, where the HSV color and the brightness represent the phase and amplitude of the pulse envelope, respectively. We then measure the transmitted pulse envelopes, as shown in Fig. 3h-k. The experimental measurements of the transmitted pulse envelope exhibit the vanishing amplitude and the whirl dislocated phase in the spatiotemporal domain corresponding to the winding number of $l = -1$. Furthermore, the phase distributions at different distances exhibit that the phase of the STVP rotates around the center along with the pulse propagation, which agrees well with the central wavelength of the pulse.

Using the experimental data and numerical $z$-propagation of the field, we calculate the real-time amplitude of the pulse and the momentum density at different instants of time, which shows the propagation evolution and diffraction of the generated STVP (Fig. 4). At the early stage of the pulse generation, Fig. 4a shows that the pressure amplitude increases as the pulse propagates along the $z$ direction, and the directions of the momentum density indicate the compression and the decompression of air. However, the zero-value amplitude is shown up when $t = -0.4ms$, which corresponds to the vortex center (Fig. 4b). Figs. 4c-d further exhibit the vortex rolls and diffract along with the propagation. Since there is an ongoing theoretical debate in community about the intrinsic OAM carried by a STVP [33,35,40], the calculation of the intrinsic OAM carried by the generated STVP is beyond the present studies, which needs more detailed theoretical and experimental investigation [SM Sec. VII].

To demonstrate the topologically robustness of STVP generation, we perturb the meta-grating structure by randomly placing fifteen photopolymer-resins particles of different shapes and sizes about 0.8cm-1cm, as shown in Fig. 3c. Moreover, as a regular perturbation to the grating, we also add one more block in the unit cell. The transmission spectrum function of the perturbed meta-grating still exhibited the same phase singularity with slightly shifted position at $\omega_0/2\pi = 7.56 \ kHz$ and $k_{x0} = 0.02k_0$ (SM Fig. S5). Adjusting the incident pulse to the perturbed central frequency, we measured the transmitted pulse envelopes at the propagation distances of 65.3mm, 80.6mm, 95.9mm and 111.2mm, respectively. Fig. 3l-o clearly show a STVP quite similar to that in the unperturbed case.

*Discussion.* In summary, our experimental results demonstrate the topologically protected generation of acoustic STVPs in two spatial dimensions and one temporal dimension, using a 1D periodic meta-grating. On the one hand, acoustic STVPs open an avenue for acoustic spacetime-structured waves, so far mostly studied in optics [41,42]. On the other hand, our new method of the generation of STVPs opens can find applications in acoustics, optics, and other types of waves. One can expect that by designing 2D metasurfaces with an additional spatial dimension, one can synthesize full-dimensional (3+1)D spatiotemporal acoustic vortices, such as vortices with arbitrarily tilted OAM [30,31] or toroidal vortices. In general, due to drastic geometric and physical differences from conventional monochromatic vortex beams, the STVPs can bring novel functionalities to acoustic/optical manipulation of particles, information transfer, and other applications [7,11,32,45-49].

Furthermore, similarly to the image processing of edge detection in the spatial domain [50-62], our STVP generator, based on the phase singularity (vortex) in the momentum-frequency domain, operates as the first-order differentiator in both spatial and temporal domains. This allows efficient extraction of the spacetime boundary information in the incident sound wavepacket (SM Sec. VIII), which can find useful applications in sonars and sensing. In our experiment, the frequency bandwidth with the near-linear dependence of the transmission amplitude near the vortex center, which provides the first-order differentiation, is about 431.7Hz.

The authors acknowledge funding through the National Key Research and Development Program of China (Grant No. 2022YFA1405200, 2022YFA1404203), the National Natural Science Foundation of China (NSFC Grants Nos. 12174340, 12174339). Z. Y. acknowledges Zhejiang Provincial Natural Science Foundation of China under Grant No. LR23A040003, and the Excellent Youth Science Foundation Project (Overseas)

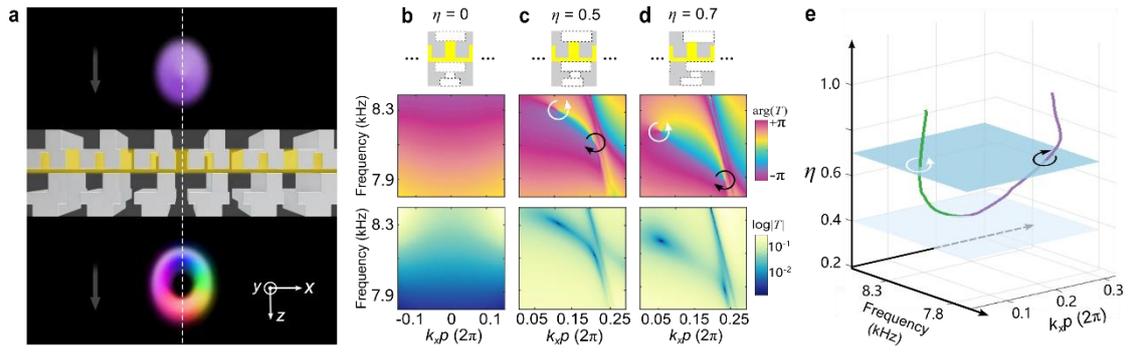

**Fig. 1| Generation of acoustic spatiotemporal vortex pulses and their topological nature.** a, Sketch of a meta-grating generating a spatiotemporal vortex pulse for airborne sound by breaking mirror symmetry, where the asymmetric structure is necessary to create the phase singularity in the spatiotemporal domain. b-d, Numerical simulations for the phase (above) and the amplitude (below) of the transmission spectrum function by breaking the mirror symmetry of the meta-grating. The synthetic parameter $\eta$ controls the deformation of symmetry breaking, and $\eta = 0, 0.5, 0.7$ correspond to b-d, respectively. The vortices with the winding numbers of +1 and -1, are indicated by white and black circles, respectively. e, Nodal lines in the extended dimension with the asymmetry parameter $\eta$, emerges at the critical values of $\eta_c = 0.40$, for the vortices with the winding number as +1 and -1 with the green and the purple lines, respectively.

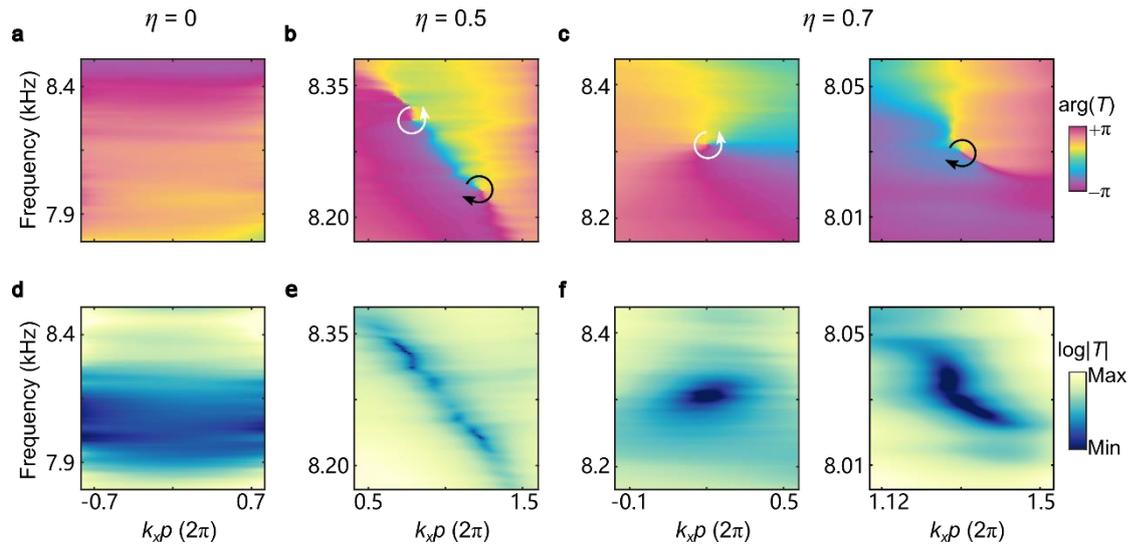

**Fig. 2| Experimental measurement for the phase (a-c) and amplitude (d-f) of transmission spectrum function for the asymmetry parameters $\eta = 0, 0.5, 0.7$, respectively.** The vortices with the winding numbers of +1 and -1, which are indicated by white and black circles, respectively.

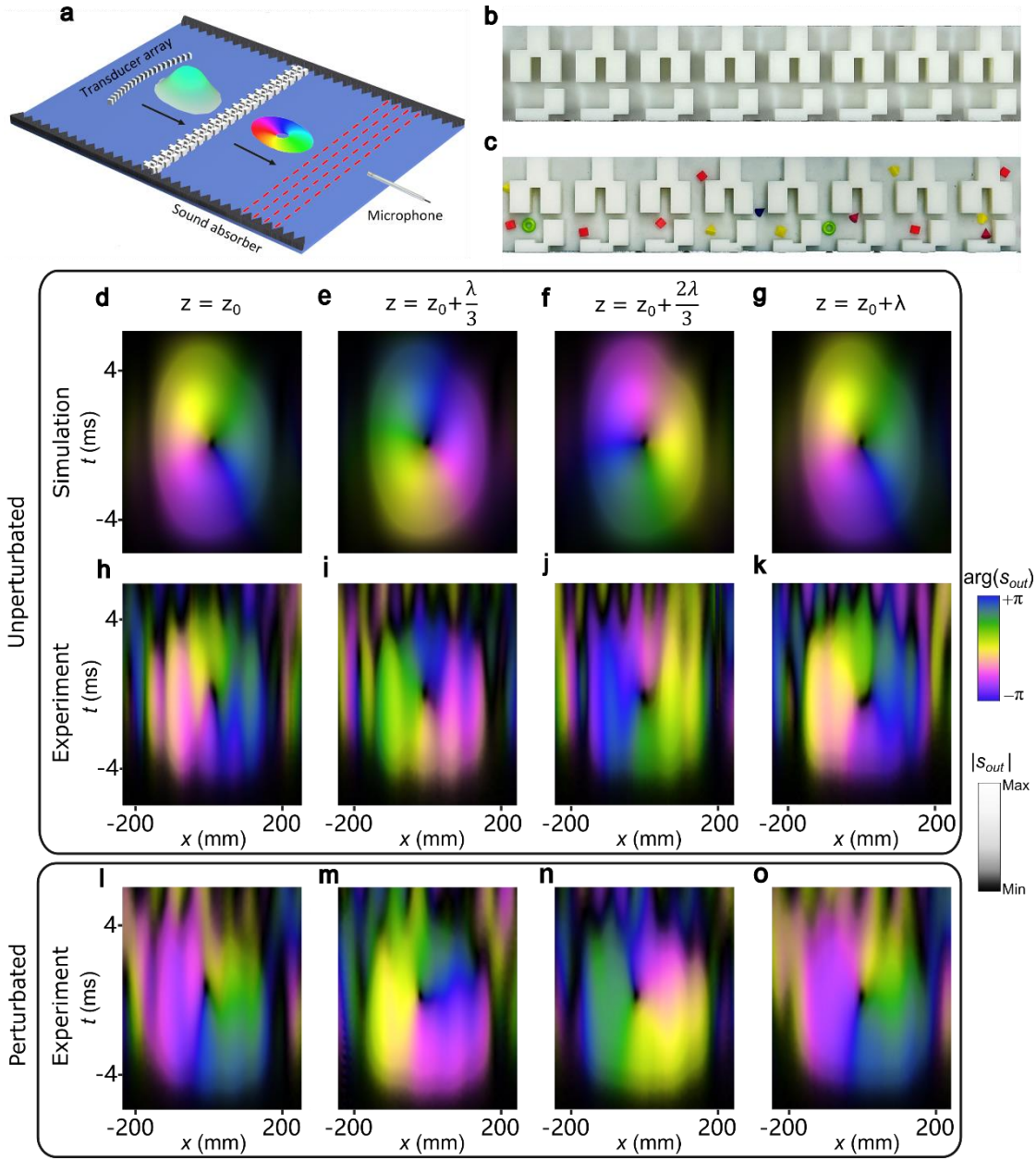

**Fig. 3| Experimental measurement of STVP generations with the acoustic meta-grating.** a, Sketch of the experimental setup with an incident Gaussian-profile pulse in both spatial and temporal domains. b, Experimental sample of the acoustic STVP generator with the asymmetry parameter $\eta = 1.0$. c, Schematics of the perturbed case where fifteen particles of different shapes such as sphere, pyramid, cube and ring are randomly placed with additional small truncations. d-g, Simulation results of the transmitted pulse envelopes at the positions which are separated by $\frac{\lambda}{3}$ from each other ($\lambda$ is the center wavelength of the pulse), respectively. h-k, Experimental measurements of the transmitted pulse envelopes at the corresponding positions of d-g, respectively. l-o, Experimental measurement of topologically protected STVP generations with perturbation, at the corresponding positions away from the sample, respectively.

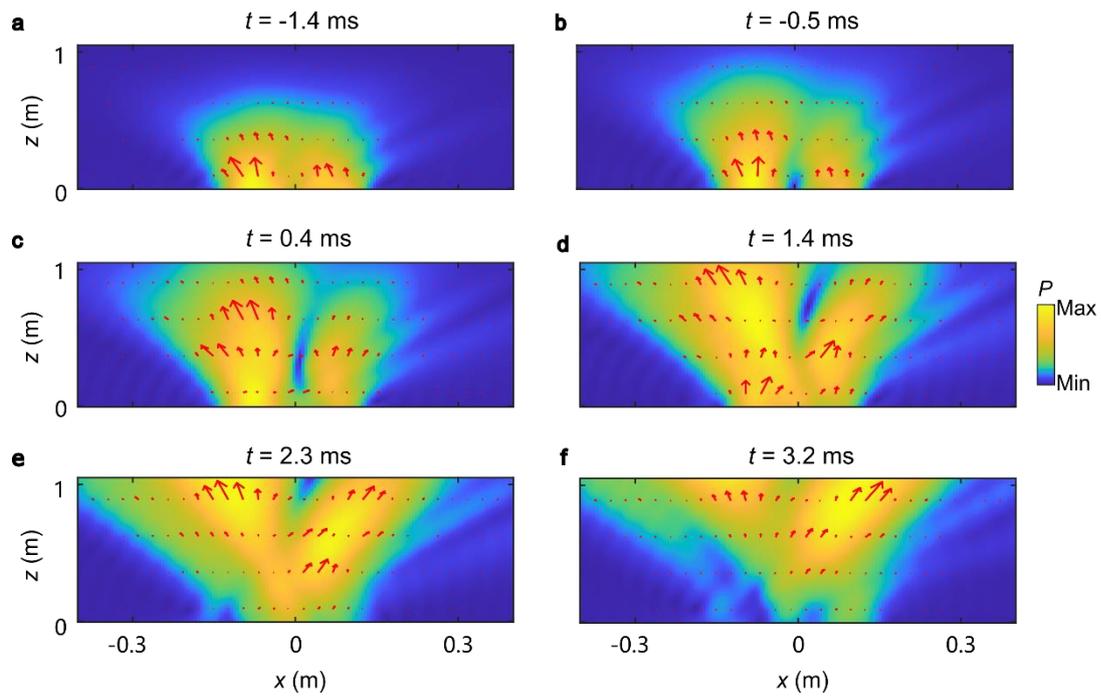

**Fig. 4| Acoustic STVP propagation in real space.** a-f, The pressure amplitude distribution (colormap) and the momentum density (arrows) of the transmitted wave are shown at the time of $-1.4, -0.5, 0.4, 1.4, 2.3, 3.2\ ms$, respectively.

# SUPPLEMENTARY MATERIALS

## Observation of acoustic spatiotemporal vortices

Hongliang Zhang[1#], Yeyang Sun[1#], Junyi Huang[1#], Bingjun Wu,
Zhaoju Yang[1*], Konstantin Y. Bliokh[2,3,4], and Zhichao Ruan[1**]

[1]*Interdisciplinary Center of Quantum Information, State Key Laboratory of Modern Optical Instrumentation, and Zhejiang Province Key Laboratory of Quantum Technology and Device, Department of Physics, Zhejiang University, Hangzhou 310027, China*
[2]*Theoretical Quantum Physics Laboratory, RIKEN Cluster for Pioneering Research, Wako-shi, Saitama 351-0198, Japan*
[3]*Centre of Excellence ENSEMBLE3 Sp. z o.o., 01-919 Warsaw, Poland*
[4]*Donostia International Physics Center (DIPC), Donostia-San Sebastián 20018, Spain*

[#] These authors equal contributions to this work
* zhaojuyang@zju.edu.cn
** zhichao@zju.edu.cn

**S1. The details of geometry parameters of acoustic spatiotemporal vortex pulse generator**

Assume that there exists a *spatiotemporal vortex pulse* (STVP) with the winding number of $l$ generated by the meta-grating. Due to the existence of mirror symmetry about the axis $x = 0$, we can easily obtain that there must exist a STVP with the winding number of $-l$ that is generated together with the STVP with the winding number of $l$. Therefore, according to the uniqueness theorem of wave equations, we can say that only the wave solution with the winding number of $l = 0$ is satisfied. As a result, without loss of generality, we suppose that the necessary condition of generating STVPs with the nonzero winding number is mirror symmetry breaking, which provides a simultaneous control for sound in both spatial and temporal domains.

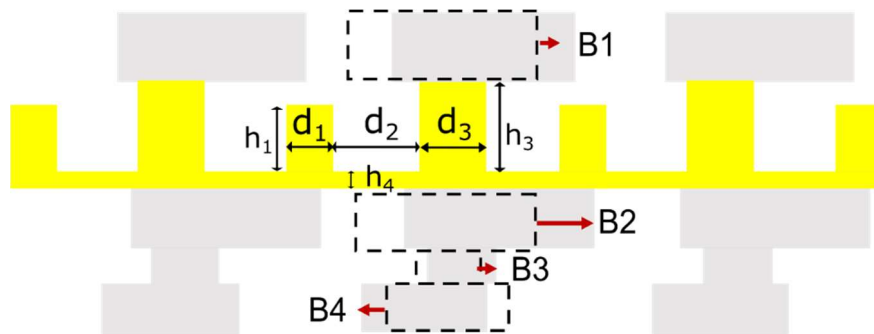

Fig. S1 The schematic of air channels in the meta-grating as shown Fig. 1a in the main text. The block marked in yellow is fixed, and the other four blocks shift toward left or right.

| Block | Length(mm) | Width(mm) | $A_i$ |
|---|---|---|---|
| B1 | 22.500 | 12.967 | 1.170 |
| B2 | 23.330 | 10.290 | 9.500 |
| B3 | 8.660 | 2.990 | 2.170 |
| B4 | 16.580 | 3.320 | -1.790 |

Table. S1 Structure parameters and shifting ratios $A_i$ for four rectangular blocks in Fig. S1. The block shift toward right (left) when $A_i$ is positive (negative).

The structure of the STVP generator consists of a fixed symmetrical air unit and four air blocks which are deformed to break mirror symmetry. The fixed part is marked in yellow in Fig. S1, and the geometric parameters are $d_1 = 5.670mm$, $d_2 = 10.000mm$, $d_3 = 7.670mm$, $h_1 = 12.997mm$, $h_3 = 16.663mm$, and $h_4 = 3.330mm$. For $\eta = 0$, the four blocks are spatially mirror-symmetrical with respect to the fixed part. When the four blocks are shifted to the left or right, the mirror symmetry of the structure is gradually broken. The offsets of the four blocks relative to the axis of symmetry are defined as: $\delta x_i = A_i \eta$, where $A_i$ is the shifting ratio different for each block and $\eta$ is a synthetic parameter characterizing the degree of asymmetry. The specific values of parameters for the four blocks and their corresponding shifting ratios are shown in Table. S1.

**S2. Experimental setup and methods to measure the transmission spectrum function**

The experimental setup is shown in Fig. S2(a). A data acquisition (Brüel & Kjær 3160-A-042-R) is used to collect the data of acoustic field and control the output waveform. Two microphones (Brüel & Kjær 4193-L-004) are connected to the data acquisition and used to measure the acoustic field. We use a power amplifier (Brüel & Kjær 2735) to amplify the input signal. The displacement platform (LINBOU NFS03) and the data acquisition are integrated into a PC. The meta-grating and the sound absorber are placed between two glasses as shown in Fig. 3(a).

To measure the transmission spectrum function in the frequency domain (as shown in Fig. S2(b)), we use 10 transducers to form a rectangle source in the spatial domain, which ensures that the spatial spectrum of the incident field is sufficiently wide but don't overlap with the non-zero diffraction order. The incident(transmitted) acoustic wave $P_{in(out)}(x,t) = s_{in(out)}(x,t)e^{-i\omega_0 t}$ is collected by two microphones, where $s_{in(out)}(x,t)$ is the pulse envelope of the input(transmitted) waves. One microphone is fixed in the acoustic field as a reference probe, and the other one probes the acoustic distribution through the displacement platform as a measurement probe. In order to obtain the transmission spectrum function of the signal, we obtain the spatial distribution of the sound field by sweeping the field, and using cross-power spectrum methods to process the data from the measurement probe and the reference probe.

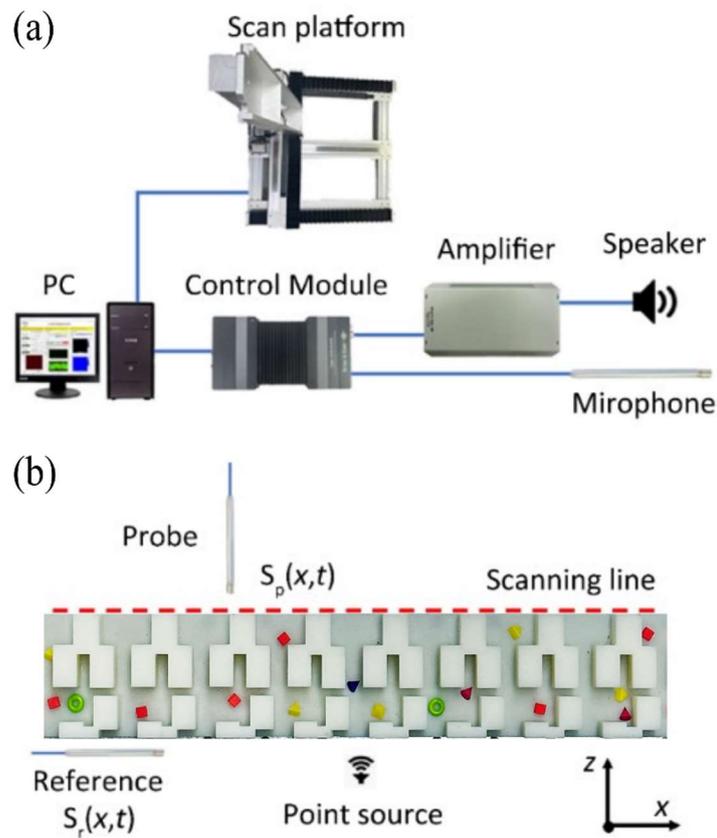

Fig. S2 The experimental setup and the process of measuring. (a) The experimental setup. (b) The process of measuring.

## S3. Diagram of the experimental structures with different shifting ratios.

Figs. S3(a-c) shows the experimental samples which fabricated with 3D printing technology with different $\eta = 0, 0.5, 0.7$, respectively.

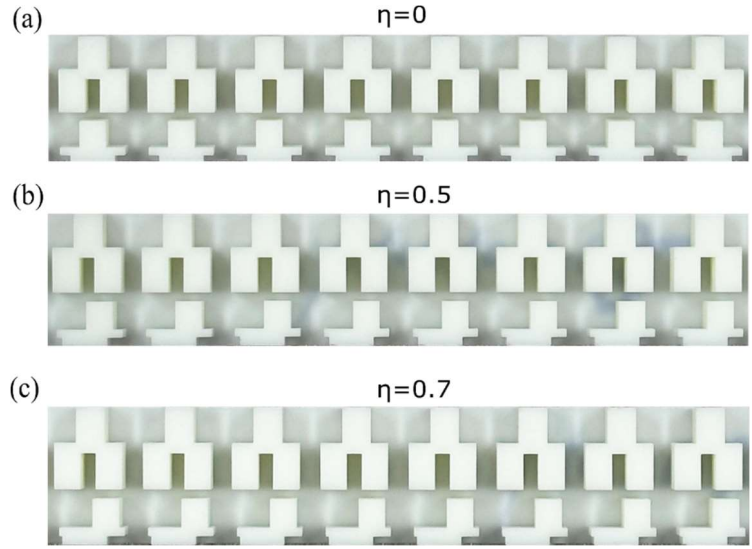

Fig. S3 (a-c) The experimental samples of the meta-gratings for $\eta = 0, 0.5, 0.7$.

## S4. The detailed derivation about STVP generation

In order to specifically depict the transformation between the incident and transmitted pulses while propagating through the STVP generator, we decompose the incident (transmitted) pulse envelope into a series of plane waves by Fourier transform $s_{in(out)}(x,t) = \iint \tilde{s}_{in(out)}(K_x, \Omega) \exp(iK_x x - i\Omega t) dK_x d\Omega$, where $K_x = k_x - k_x^0$ is the wavevector component shifted along x direction, and $\Omega = \omega - \omega_0$ is the sideband angular frequency from the center one $\omega_0$. In the polar coordinate, the expression is converted to $s_{in(out)}(r, \theta) = \int_0^\infty \int_0^{2\pi} \tilde{s}_{in(out)}(\rho, \chi) \exp[i\rho r \cos(\chi + \theta)] \rho d\rho d\chi$ and where $r = \sqrt{x^2 + t^2}$, $\theta = \tan^{-1}(t/x)$, $\rho = \sqrt{K_x^2 + \Omega^2}$, $\chi = \tan^{-1}(\Omega/K_x)$. According to the restriction of the winding number around vortex, within the small enough region, the amplitude of transmission spectrum function $T$ for the STVP generator can be written as $f(\rho)$, and without loss of general $T$ can be expressed as

$$T(\rho, \chi) = f(\rho)(Ae^{il\chi} + Be^{-il\chi}),$$

where $l$ is the winding number of the vortex, $A$ and $B$ are the two constant parameters, respectively. We note that when the phase singularity is anisotropic, both $A$ and $B$ are nonzero. The overall topological charge is $l$ if $|A|$ is larger than $|B|$, and $-l$ otherwise. Suppose that an incident pulse with zero topological charge $\tilde{s}_{in}(\rho)$ impinges on the meta-gratings, the transmitted pulse can be expressed as:

$$s_{out}(r,\theta) = \iint \tilde{s}_{in}(\rho) f(\rho) \left( A e^{il\chi} + B e^{-il\chi} \right) \cdot e^{i(k_x x - \omega t)} dk_x d\omega$$

$$= \int_0^\infty \tilde{s}_{in}(\rho) f(\rho) \rho \int_0^{2\pi} A e^{il\chi} e^{i(\rho\cos\chi \cdot r\cos\theta - \rho\sin\chi \cdot r\sin\theta)} + B e^{-il\chi} e^{i(\rho\cos\chi \cdot r\cos\theta - \rho\sin\chi \cdot r\sin\theta)} d\chi d\rho$$

$$= \int_0^\infty \tilde{s}_{in}(\rho) f(\rho) \rho \left[ A e^{-il\theta} \int_0^{2\pi} e^{i[-l(-\chi-\theta) + \rho r\cos(\chi+\theta)]} d\chi + B e^{il\theta} \int_0^{2\pi} e^{i[-l(\chi+\theta) + \rho r\cos(\chi+\theta)]} d\chi \right] d\rho.$$

Taking $\eta = -\chi - \theta + \frac{\pi}{2}$ and $\tau = \chi + \theta + \frac{\pi}{2}$, then $d\eta = -d\chi$ and $d\tau = d\chi$, by periodicity, the integral of $d\chi$ becomes:

$$A e^{-il\theta} \int_0^{2\pi} e^{i[-l(-\chi-\theta) + \rho r\cos(\chi+\theta)]} d\chi + B e^{il\theta} \int_0^{2\pi} e^{i[-l(\chi+\theta) + \rho r\cos(\chi+\theta)]} d\chi$$

$$= -A e^{-il\theta} \int_{-\theta+\frac{\pi}{2}}^{-2\pi-\theta+\frac{\pi}{2}} e^{i[-l\left(\eta-\frac{\pi}{2}\right) + \rho r\cos(\frac{\pi}{2}-\eta)]} d\eta + B e^{il\theta} \int_{\theta+\frac{\pi}{2}}^{2\pi+\theta+\frac{\pi}{2}} e^{i[-l\left(\tau-\frac{\pi}{2}\right) + \rho r\cos(\tau-\frac{\pi}{2})]} d\tau$$

$$= -i^l A e^{-il\theta} \int_0^{-2\pi} e^{i[-l\eta + \rho r\sin(\eta)]} d\eta + i^l B e^{il\theta} \int_0^{2\pi} e^{i[-l\tau + \rho r\sin(\tau)]} d\tau$$

$$= i^l A e^{-il\theta} \int_0^{2\pi} e^{i[-l\eta + \rho r\sin(\eta)]} d\eta + i^l B e^{il\theta} \int_0^{2\pi} e^{i[-l\tau + \rho r\sin(\tau)]} d\tau$$

$$= 2\pi i^l (A e^{-il\theta} + B e^{il\theta}) J_l(\rho r)$$

Where $J_l(\rho r) = \frac{1}{2\pi} \int_0^{2\pi} e^{i[-l\gamma + \rho r\sin(\gamma)]} d\gamma$ is the *l*-th order Bessel function of the first kind. Therefore, the transmitted pulse envelope can be expressed as $s_{out}(r,\theta) = 2\pi i^l (A e^{-il\theta} + B e^{il\theta}) \int_0^\infty \tilde{s}_{in}(\rho) f(\rho) J_l(\rho r) \rho d\rho$. Therefore, the transmitted wave packet can be determined to be a STVP with a winding number of $-l$, which is opposite to that of the vortex in the $k_x - \omega$ domain.

## S5. Transmission spectrum function of the structure with synthetic parameter $\eta = 1.0$.

We measure the transmission spectrum function $T$ of the unperturbed meta-grating with synthetic parameter $\eta = 1.0$ in the experiment and Fig. S4(a, b) show the amplitude and phase distribution of the transmission spectrum function with respect to $K_x = k_x - k_x^0$ and $\Omega = \omega - \omega_0$, respectively. The phase distribution of $T$ exhibits that the vortex in the $k_x - \omega$ domain still survives, which leads to a zero amplitude at $\omega_0/2\pi = 8.02 \times 10^3 Hz$ and $k_{0x} = 0.01 k_0$. Furthermore, the blue and orange circles in Fig. S4(c, d) correspond to the amplitudes and phases of the spectrum function along $K_x/k_0 = 0$ and $\Omega/\omega_0 = 0$ respectively.

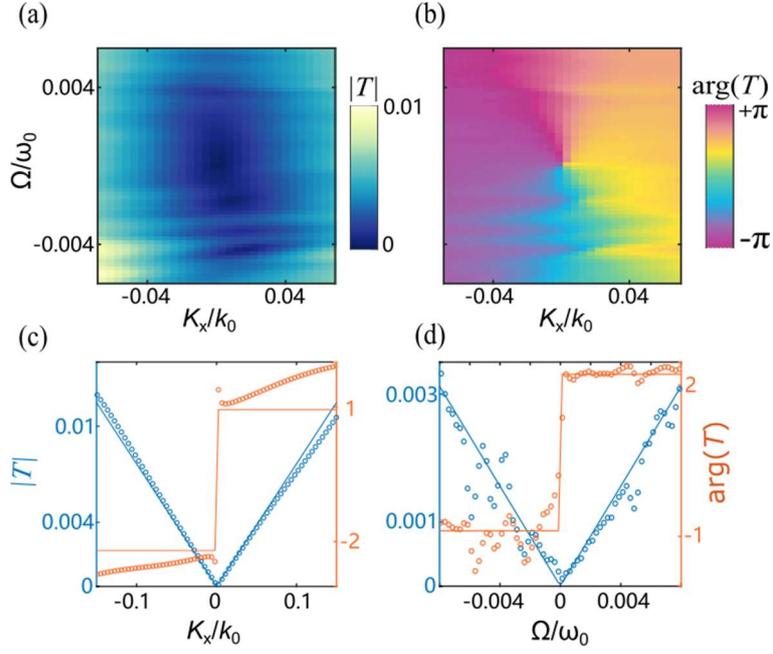

Fig. S4 Transmission spectrum function of the unperturbed meta-grating with synthetic parameter $\eta = 1.0$ as Fig. 3b. a) Amplitude and b) phase distributions of the spectrum function with respect to $K_x$ and $\Omega$. The amplitudes and phases of the spectrum function along (c) $K_x = 0$ and (d) $\Omega = 0$. The blue circles and lines correspond to the amplitudes of simulation results and the fitting ones with Eq. (S1), and the orange rhombi and lines correspond to the phases, respectively. $\omega_0$ and $k_x$ are the frequency and the wavenumber of light in vacuum at the center wavelength $\lambda_0$.

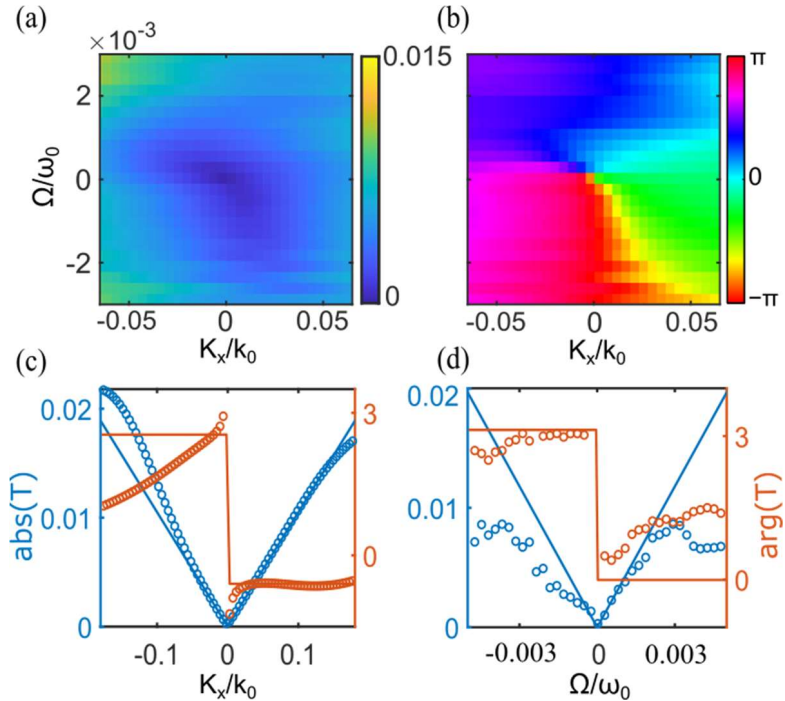

Fig. S5 Transmission spectrum function of the perturbed meta-grating of Fig. 3c, with a block in each unit cell and the randomly placed fifteen particles. a) Amplitude and b) phase distributions of the spectrum function with respect to $K_x$ and $\Omega$. The amplitudes and phases of the spectrum function along (c) $K_x = 0$ and (d) $\Omega = 0$. The blue circles and lines correspond to the amplitudes of simulation results and the fitting ones with Eq. (S1), and the orange rhombi and lines correspond to the phases, respectively. $\omega_0$ and $k_0$ are the frequency and the wavenumber of light in vacuum at the center wavelength $\lambda_0$.

We also measure the transmission spectrum function $T$ of the perturbed spatiotemporal differentiator which has a block in each unit cell and the randomly placed fifteen particles (Fig. 3c). Similarly, Fig. S5(a, b) show the amplitude and phase distribution of the transmission spectrum function with respect to $K_x$ and $\Omega$, respectively. The phase distribution of $T$ exhibits that the vortex in the $k_x - \omega$ domain still survives even with the perturbations, which leads to a zero amplitude at $\omega_0/2\pi = 7.56 \times 10^3 \text{Hz}$ and $k_{0x} = 0.02k_0$ . The blue and red circles in Fig. S5(c, d) correspond to the amplitudes and phases of the spectrum function along $K_x/k_0 = 0$ and $\Omega/\omega_0 = 0$ respectively.

## S6. Experimental setup and measurement principle of STVP.

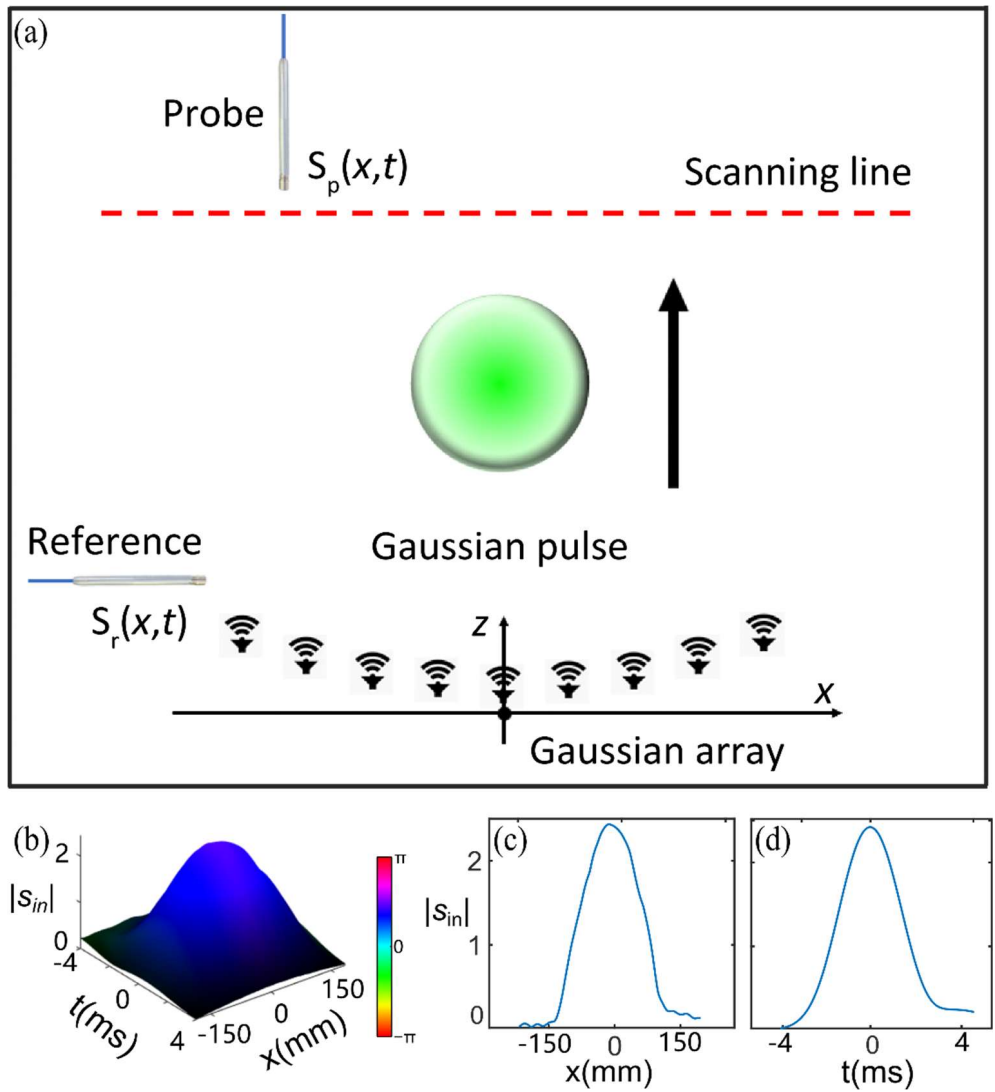

Fig. S6 Experimental setup and the incident pulse which have the Gaussian-like profiles in both the spatial and temporal domains. (a) Experimental setup for measuring STVP. (b) Amplitude and phase distribution of the envelope of the incident pulse. Amplitude distribution of the envelope along (c) $t = 0$ and (d) $x = 0$.

We use a curved transducer array to simultaneously generate a series of pulses with spatiotemporal Gaussian envelope at the center frequency $\omega_0/2\pi = 7.56 \times 10^3 Hz$ as shown in Fig. S6(a). The distribution of curved transducers array satisfies Gaussian function distribution ($z = \exp(x^2) - 1$), and the gap between the transducers in x direction is 1cm. We put the sample 50cm away from the transducers array. The distribution of $s_{in}(x,t)$ is shown in Fig. S6(b). The height shows the amplitude and the color indicates the phase distribution, respectively. Figures S6(c, d) show the amplitude distribution of the envelope along (c)$t = 0$ and (d)$x = 0$.

## S7. Discussion about the calculation of the intrinsic transverse OAM

We note that there is an ongoing theoretical debate about the value of transverse OAM carried by STVPs in Ref. [33, 35, 40]. Depending on the definition of the intrinsic OAM with respect to the photon/phonon probability or energy centroid, this intrinsic OAM (normalized per photon/phonon) equals $L_y^{int} = l\frac{\gamma+\gamma^{-1}}{2}$ in Ref. [33] and $L_y^{int} = l\frac{\gamma}{2}$ in Ref. [40], where $\gamma$ is the ellipticity of the STVP shape in the (z,x) plane. For $\gamma \simeq 7.6$ for the STVP in our experiment, two theoretical values for the intrinsic OAM are about 3.9 and 3.8, respectively. Imperfections of the STVP intensity profile as compared to the ideal elliptical shape make little difference in this case.

## S8. The simulation of detecting the sharp changes of pulse envelopes.

To demonstrate spatiotemporal differentiator, we consider the perturbed meta-grating as shown in Fig. 3c, where the corresponding transmission spectrum function $T$ is shown in Fig. S5. The phase distribution of $T$ exhibits that the vortex in the $k_x - \omega$ domain still survives, which leads to a zero amplitude at $\omega_0/2\pi = 7.56 \times 10^3 Hz$ and $k_{0x} = 0.02k_0$. Furthermore, $T$ exhibit a good linear dependence on $K_x/k_0 = 0$ and $\Omega/\omega_0 = 0$ within a certain bandwidth, and the phase shifts with $\pi$ at the minima occurring at $K_x = 0$ and $\Omega = 0$, respectively, which indicates that the structure still enables the first-order differentiation in both the spatial and temporal domains. Around $K_x = 0$ and $\Omega = 0$, the transmission spectrum function $T$ has the form:

$$T = C_x K_x + C_t \Omega \qquad (S1),$$

where $C_x = 0.105\, exp(-0.598i)/k_0$ and $C_t = 3.93\, exp(-0.012i)/\omega_0$ are two complex numbers. Here, $\omega_0$ is the frequency at the center wavelength $\lambda_0$. The results indicate that the structure is a spatiotemporal differentiator, which can be applied to detect sharp changes of pulse envelopes, similar to the image processing of edge detection by the spatial differentiators in the real space (Ref. [32, 50, 55, 60]).

To demonstrate the spatiotemporal boundary extraction, we simulate the incident pulse with the amplitude modulation as a submarine in the spatiotemporal domain in Fig. S7(a), and the phases of the incident pulse envelope are binary with only 0 or $\pi$ as Fig. S7(b) thus without phase singularities. We simulate the pulse transmitted through the perturbed meta-grating with the central frequency at $\omega_0/2\pi = 7.56 \times 10^3 Hz$ and central transverse wavevector component $0.02k_0$. Fig. S7(c) and S7(d) correspond to the amplitude and phase distributions of the transmitted pulse envelope, respectively.

Moreover, the phase distribution of the transmitted pulse exhibits the generation of large numbers of adjacent STVPs in the spatiotemporal domain. As the generated STVPs interfere with each other, the constructive interference occurs at the sharp changes of the incident pulse envelope in both spatial and temporal domains, while the destructive one strongly takes place where the amplitudes have slight variations.

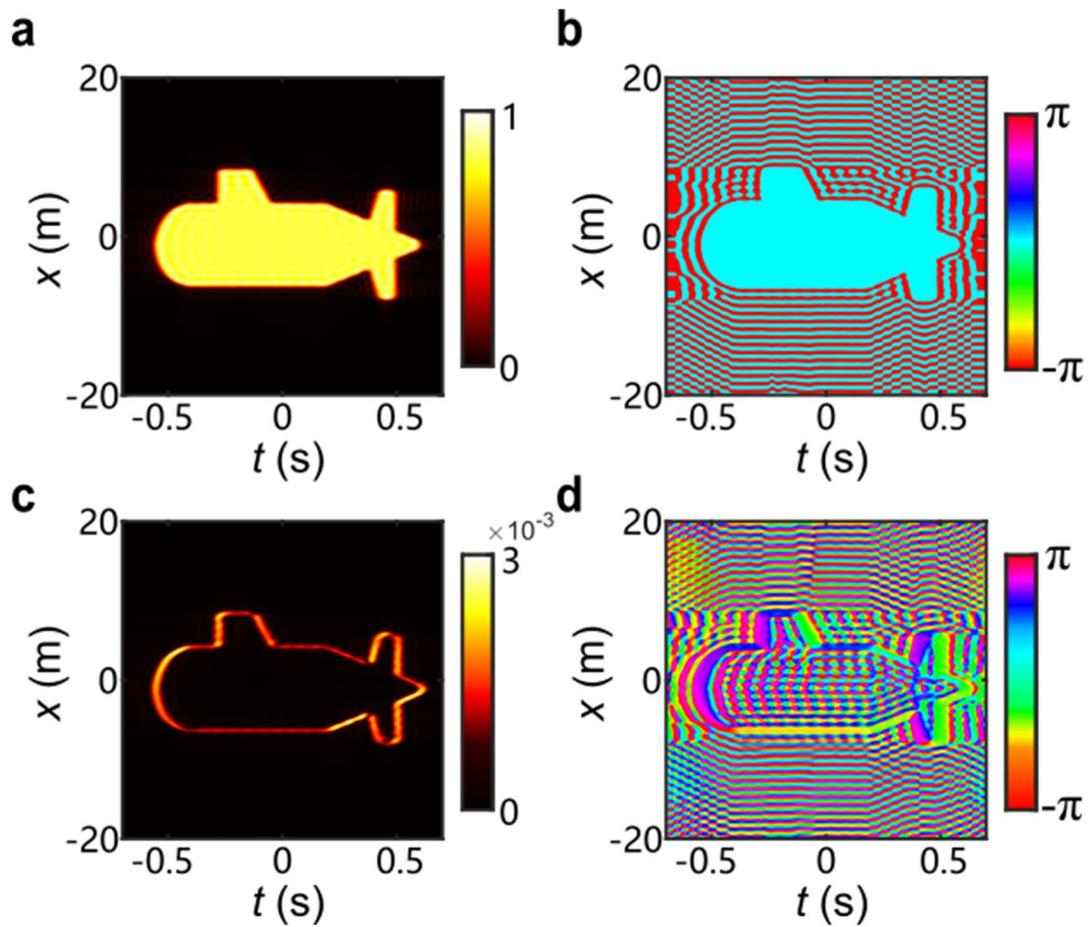

Fig. S7 Spatiotemporal boundary extraction for an arbitrary amplitude modulated spatiotemporal pulse. (a) Amplitude and (b) phase distributions of an incident pulse envelope as a torpedo. The phases of the incident pulse envelope are binary with only 0 or $\pi$, without phase singularities. (c) Amplitude and (d) phase distributions of the transmitted pulse envelope.